\documentclass[twocolumn,superscriptaddress,letter]{revtex4}%
\usepackage{amssymb}
\usepackage{color}
\usepackage{graphicx}
\usepackage{dcolumn}
\usepackage{bm}
\usepackage[header,title,page,titletoc]{appendix}
\usepackage{amsmath}
\usepackage{amsfonts}%
\providecommand{\U}[1]{\protect \rule{.1in}{.1in}}
\begin{document}
\title{Hierarchical Topological Superconductor -- a Majorana Vortex Lattice Model}
\author{Jiang Zhou}
\author{Ya-Jie Wu}
\author{Rong-Wu Li}
\author{Jing He}
\author{Su-Peng Kou \thanks{Corresponding author Email: spkou@bnu.edu.cn}}
\affiliation{Department of Physics, Beijing Normal University, Beijing 100875, PR China}
\date{November 16,2012}

\begin{abstract}
In this paper we study an $s$-wave topological superconductor (SC) with a
square vortex-lattice. We proposed a topological Majorana lattice model to
describe this topological state which was supported by the numerical
calculations. We found that the Majorana lattice model is really a
"topological SC" on the parent topological SC. Such hierarchy structure
becomes a new holographic feature of the topological state.

\end{abstract}
\maketitle

%%%%%%%%%%%%%%%%%%%%%%%%%%%%%%%%%

\section{Introduction}

Recently, the search for exotic states supporting Majorana fermions (modes)
has attracted increasing interests due to their potential applications in
fault-tolerant quantum
computations\cite{YT,SDS,LFu,NR,MS,JDS,RML,DAI,IPR,WBI,JAL,CNA,SDAS,STE,AKI}.
A creative proposal is the proximity effect between $s$-wave superconductor
(SC) and topological insulator\cite{LFu}. This system exhibits non-trivial
topological properties, including the nontrivial Chern number in the momentum
space, the chiral Majorana edge states, in particular, the Majorana mode
around the $\pi$-flux vortex. Another possible example of such quantum exotic
states is the chiral $p+ip$ topological superconductors\cite{NR}. The
quantized magnetic vortex ($\pi$-flux) in the chiral $p+ip$ topological SC
hosts the Majorana zero modes and obeys non-Abelian statistics. When there are
two $\pi$-fluxes nearby, the intervortex quantum tunneling effect occurs and
the Majorana modes on two $\pi$-fluxes couple. The tunneling amplitude is
determined by the overlap of the wave function of the localized Majorana bound
states\cite{MCH2,MCH}. Thus, such tunneling must be taken into account when
the average distance between localized $\pi$-fluxes becomes the order of the
Majorana bound state decay length. For the topological quantum computation
based on the non-Abelian anyons, the tunneling effect would split the
zero-energy bound states and lift the ground state degeneracy. Beside, the
sign of energy splitting is important for understanding the many-body
collective states\cite{CNA,CG,AF}.

However, till now people have not identify the experimental realizations of
$p$-wave SC in condensed matter physics\cite{DAI,ASTE}. Instead, people
pointed out that the chiral $p$-wave SC may be realized in ultra-cold
fermionic atom systems\cite{VGU,TMI,CZH,NRC,YNS}. Theoretically, $p$-wave SC
may be realized via a p-wave Feshbach resonance in experiment. Due to huge
particle-loss\cite{STE}, this idea also has not been realized. For this
reason, people proposed a more advantageous scenario based on $s$-wave SC of
cold fermionic atoms with laser-field-generated effective spin-orbit (SO)
interactions and a large Zeeman field\cite{MS}. In this scenario, there is a
non-Abelian topological phase that is different from $p$-wave SC, in which the
SO interaction plays the role of the $p$-wave SC order parameter.

In this paper, we start from this $s$-wave topological SC model and show that
there exists the Majorana mode hosted by the $\pi$-flux (quantized magnetic
vortex). Then we focus on the super-lattice of $\pi$-fluxes. Because each
$\pi$-flux traps a Majorana mode, we can have a lattice model of the Majorana
modes that is called the Majorana lattice model. We took into account of zero
mode tunneling that couples the vortex sites. We found that this model shows
nontrivial topological properties, including a nonvanishing Chern number,
chiral Majorana edge state\cite{VILL}. In this sense, the Majorana lattice
model is really a "topological SC" on the parent topological SC. Such
hierarchy relationship between the Majorana lattice model and the $s$-wave
topological SC model with vortex-lattice is a new holographic feature of the
topological states.

The paper is organized as follow: We first introduce the topological $s$-wave
SC model on a square lattice in Sec. II, then study the topological properties
of this model. In Sec. III, the Majorana mode around a $\pi$-flux is obtained
and the intervortex quantum tunneling effect is also studied. We also study
the topological SC with a square vortex-lattice numerically. In Sec. IV we
write down a Majorana lattice model to describe the coupling effect between
Majorana modes trapped in vortices, and the topological properties of this
Majorana lattice model are also analyzed. Finally we draw the conclusion in
Sec. V.

\section{The s-wave pairing topological superconductor with Rashba
spin-orbital coupling}

As a starting point, the $s$-wave pairing SC with Rashba SO coupling is
defined on a square lattice\cite{MS}, which is described by
\begin{equation}
\mathcal{H}=\mathcal{H}_{k}+\mathcal{H}_{so}+\mathcal{H}_{sc}%
\end{equation}
where the kinetic term $\mathcal{H}_{k}$, the Rashba spin-orbital (SO)
coupling term $\mathcal{H}_{so}$, and the superconducting pairing term
$\mathcal{H}_{sc}$ are given as
\begin{equation}%
\begin{split}
\mathcal{H}_{k}=  &  -t_{s}\sum_{j\sigma}\sum_{\mu=\vec{x},\vec{y}}%
(c_{j+\mu \sigma}^{\dag}c_{j\sigma}+c_{j-\mu \sigma}^{\dag}c_{j\sigma})\\
&  -u\sum_{j\sigma}c_{j\sigma}^{\dag}c_{j\sigma}-h\sum_{j\sigma}c_{j\sigma
}^{\dag}\sigma^{z}c_{j\sigma},\\
\mathcal{H}_{so}=  &  -\lambda \sum_{j}[(c_{j-\vec{x}\downarrow}^{\dag
}c_{j\uparrow}-c_{j+\vec{x}\downarrow}^{\dag}c_{j\uparrow})\\
&  +i(c_{j-\vec{y}\downarrow}^{\dag}c_{j\uparrow}-c_{j+\vec{y}\downarrow
}^{\dag}c_{j\uparrow})]+H.c,\\
\mathcal{H}_{sc}=  &  -\Delta \sum_{j}(c_{j\uparrow}^{\dag}c_{j\downarrow
}^{\dag}+H.c)
\end{split}
\end{equation}
where $c_{j\sigma}$ ($c_{j\sigma}^{\dag}$) annihilates (creates) a fermion at
site $j=(j_{x},j_{y})$ with spin $\sigma=(\uparrow,\downarrow)$, $\mu=\vec{x}$
or $\vec{y}$, which is a basic vector for the square lattice. $\lambda$ serves
as the SO coupling constant and $\Delta$ as the SC pairing order parameter. In
addition, in order to observe nontrivial phases supporting Majorana modes
excitation, the chemical potential $u$ and the Zeeman term $h$ also be
included in this model.

The Hamiltonian $\mathcal{H}$ can be transformed into momentum space from
$c_{j}=1/\sqrt{L}\sum_{k}c_{k}e^{-ikR_{j}}$. Writing $c_{j}$ in the
particle-hole basis $\psi_{k}^{\dag}=(c_{k\uparrow},c_{k\downarrow
},c_{-k\uparrow}^{\dag},c_{-k\downarrow}^{\dag})$, we obtain its Bogoliubov-de
Gennes(BDG) form
\begin{equation}
\mathcal{H}=\sum_{BZ}\psi_{k}^{\dag}\mathcal{H}(k)\psi_{k}d^{2}k,
\end{equation}
where the Bloch Hamiltonian $\mathcal{H}(k)$ is a $4\times4$ matrix
\begin{equation}
\mathcal{H}(k)=\left(
\begin{array}
[c]{cccc}%
\epsilon(k)-h & g(k) & 0 & -\Delta \\
g^{\ast}(k) & \epsilon(k)+h & \Delta & 0\\
0 & \Delta & -\epsilon(k)+h & g^{\ast}(k)\\
-\Delta & 0 & g(k) & -\epsilon(k)-h
\end{array}
\right)
\end{equation}
with
\begin{equation}%
\begin{split}
\epsilon(k)  &  =-2t_{s}(\cos k_{x}+\cos k_{y})-u\\
g(k)  &  =-2\lambda(\sin k_{y}+i\sin k_{x}).
\end{split}
\end{equation}
Then the energy spectrum $\epsilon_{\pm}$ is obtained by diagonalizing
$\mathcal{H}(k)$ as
\begin{equation}
\epsilon_{\pm}=\pm \lbrack m^{2}(k)+|g(k)|^{2}+h^{2}\pm2\sqrt{|g(k)|^{2}%
\epsilon^{2}(k)+h^{2}m^{2}(k)}]^{\frac{1}{2}}%
\end{equation}
where $m(k)=\sqrt{\epsilon^{2}(k)+\Delta^{2}}$.

\begin{figure}[ptb]
\scalebox{0.36}{\includegraphics* [0.3in,0in][10in,7.2in]{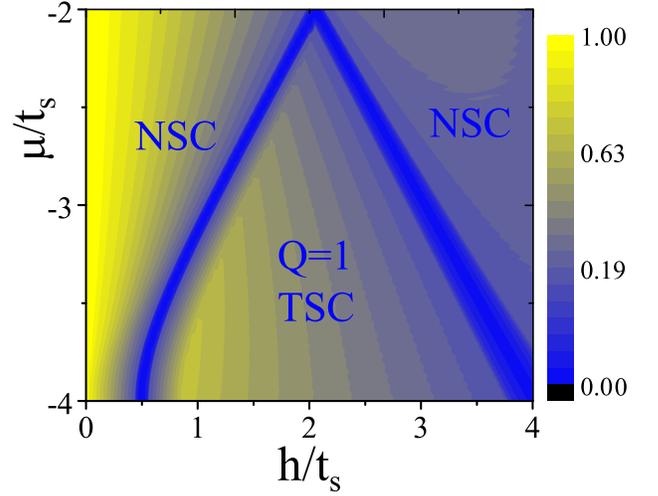}}
\caption{(Color online) The phase diagram ($\lambda=\Delta=0.5t_{s}$): The
blue lines are the boundaries between the $\mathcal{Q}=1$ topological SC (TSC)
and the normal SC (NSC) with trivial topological properties. The color denotes
the size of the energy gap.}%
\label{phase}%
\end{figure}

\begin{figure}[h]
\scalebox{0.35}{\includegraphics[width = 25.0cm]{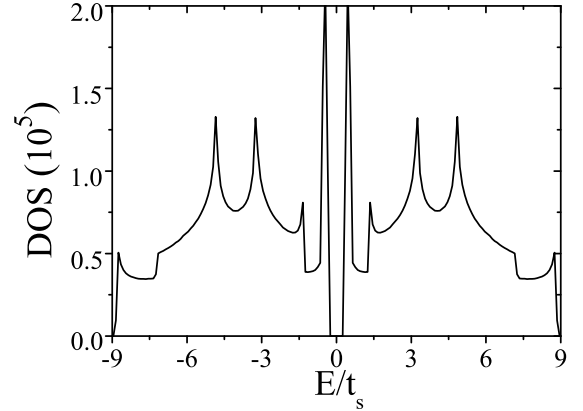}} \caption{(Color
online) The density of states (DOS) of the $s$-wave SC model for the case of
$u=-4t_{s}$, $\lambda=0.5t_{s}$, $\Delta=0.5t_{s}$, and $h=0.8t_{s}$. There
exists an energy gap $\Delta E=0.7t_{s}$.}%
\label{dosa}%
\end{figure}

In Fig.\ref{phase}, we plot the global phase diagram. The blue lines are the
boundaries between the topological SC and the SC with trivial topological
properties. In the topological SC, the ground state has a nontrivial
topological invariant (Chern-number) $\mathcal{Q}=1$\cite{MS}. Such
topological invariant is robust, for the Rashba SO coupling can be mapped into
a $p+ip$ SC gap through a suitable unitary transformation (see details in
Ref.\cite{MS}). We also study the density of state (DOS) in the topological
superconductor numerically and show the results in Fig.\ref{dosa}. From this
result one can see that there exists a finite energy gap of the topological SC
state. In other regions of the phase diagram, the ground states are the SC
state with trivial topological properties.

\begin{figure}[ptbh]
\includegraphics[width = 10.0cm]{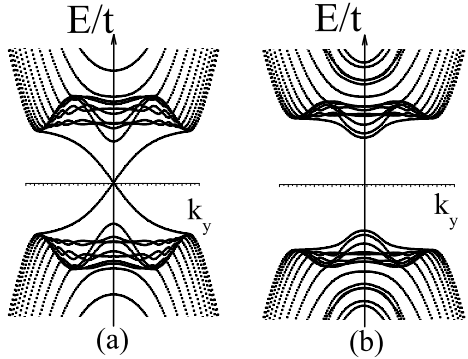}\caption{(Color online) The spectral
flow of Hamiltonian $H$ on cylinder with $u=-4t_{s}$, $\lambda=0.5t_{s}$,
$\Delta=0.5t_{s}$, and $h=0.8t_{s}$ for (a) but $h=0.4t_{s}$ for (b). For
topological SC, there exist chiral gapless edge states in (a); while for
normal SC there is no edge state in (b).}%
\label{6}%
\end{figure}

The operators of the Majorana edge states of this topological SC as shown in
Fig.3 satisfy $\gamma_{0}^{\dag}=\gamma_{0}$ , which indicates the edge states
are really Majorana fermions. Indeed, we can see that the particle-hole
operator $\mathcal{C}=\sigma^{x}\otimes \mathbf{1}$ acts on $\mathcal{H}(k)$
as
\begin{equation}
\mathcal{C}\mathcal{H}(k)\mathcal{C}^{-1}=-\mathcal{H}^{\ast}(-k)
\end{equation}
implies that the Bogoliubov quasi-particle operator follows $\gamma_{-k}%
^{\dag}=\gamma_{k}$ . From Fig.3 (a), one can find that the Majorana edge
modes cross zero energy at $k_{y}=0$ for the topological SC with $u=-4t_{s}$,
$\lambda=0.5t_{s}$, $h=0.8t_{s}$. The odd number of crossings leads to the
topological protected Majorana zero modes. For the trivial SC with $u=-4t_{s}%
$, $\lambda=0.5t_{s}$, $h=0.4t_{s},$ there is no such crossing (Fig.3 (b)).

\section{Majorana zero modes around vortices of the topological SC}

\subsection{Majorana zero modes around a pair of vortices}

\begin{figure}[ptb]
\scalebox{0.43}{\includegraphics* [1.5in,1in][9in,7.2in]{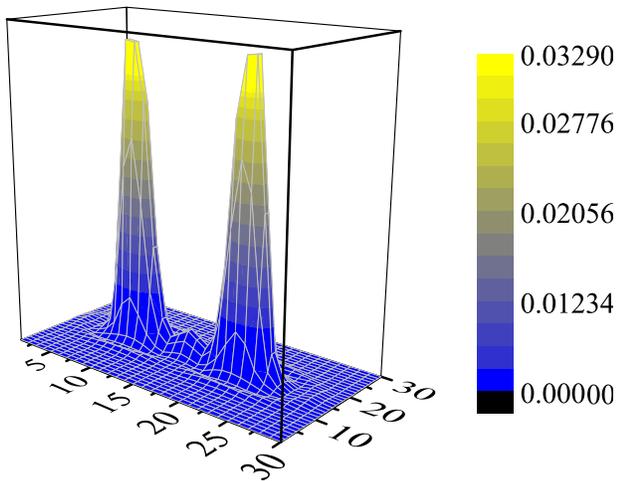}}
\caption{(Color online) The particle density distribution of two localized
zero mode states around the two $\pi$-flux-vortices for the case of
$u=-4t_{s},$ $\lambda=0.5t_{s},$ $h=0.8t_{s},$ $\Delta=0.5t_{s}.$}%
\label{pden}%
\end{figure}\begin{figure}[h]
\scalebox{0.35}{\includegraphics* [0.5in,0.5in][10in,7.5in]{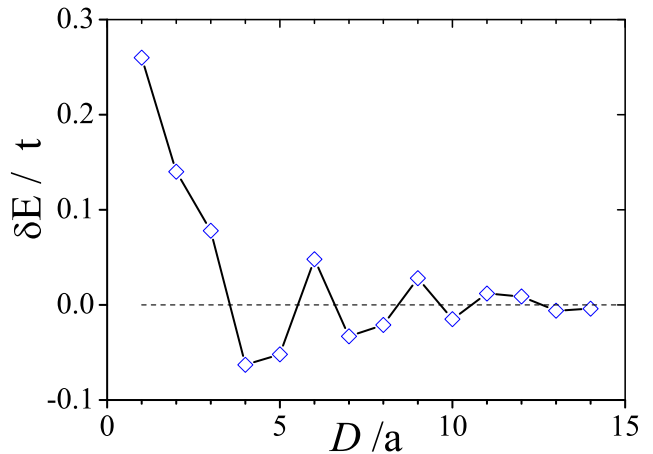}}
\caption{(Color online) The energy splitting $\delta E$ as the function of the
spacial distance of two $\pi$-flux-vortices, decays exponentially and shows
the oscillating behavior with $D$. The parameters are chosen as $u=-4t_{s},$
$\lambda=0.5t_{s},$ $h=0.8t_{s},$ $\Delta=0.5t_{s}$. $a$ is the lattice
constant.}%
\label{spl}%
\end{figure}

We now start with the discussion on the Majorana zero modes which associate
with a pair of $\pi$-flux-vortices. The particle density distribution of the
electrons is shown in Fig.\ref{pden}. The main result shows that there appear
two approximately zero modes in the presence of two well-separated $\pi
$-flux-vortices. When two $\pi$-fluxes are well separated, the quantum
tunneling effect can be ignored and we have two quantum states with zero
energy. On the other hand, for the small spatial distance $D$, the vortices
interaction becomes stronger and the energy splitting can not be neglected. As
shown in Fig.\ref{spl}, the energy splitting $\delta E$ as the function of
$D$, oscillates and decreases exponentially. Likewise, a pair of $\pi
$-flux-vortices in the topological $p$-wave SC or quasi-hole in the Moore-Read
state has the qualitative similar behavior\cite{MCH,MCH2,MBA}.

\subsection{Topological properties of the topological SC with a square
vortex-lattice}

\begin{figure}[h]
\scalebox{0.34}{\includegraphics* [0.5in,0.5in][10in,7.4in]{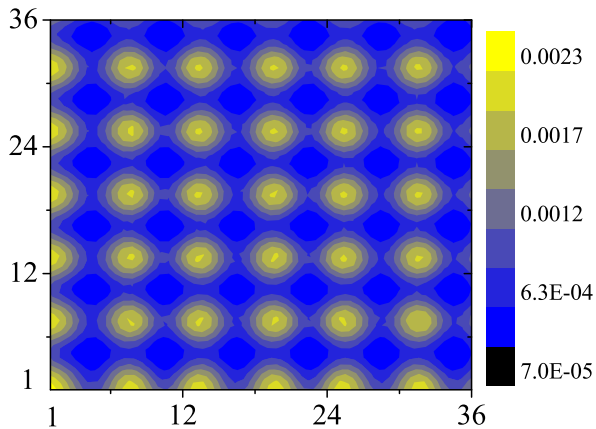}}
\caption{(Color online) The illustration of vortex-lattice: the distance
between two nearest vortices is $6a$. The parameters are chosen as
$u=-4t_{s},$ $\lambda=0.5t_{s},$ $h=0.8t_{s},$ $\Delta=0.5t_{s}$. Each vortex
traps a Majorana mode. Thus, in Sec.IV we have a tight-binding Majorana
lattice model to describe the low energy physics of the multi-Majorana-mode.}%
\label{vlat}%
\end{figure}

Next, we study the topological SC with a square vortex-lattice numerically.
The illustration of the vortex-lattice ($\pi$-flux-lattice) was shown in
Fig.\ref{vlat}. \begin{figure}[ptb]
\scalebox{0.33}{\includegraphics* [0.4in,0in][11in,7.4in]{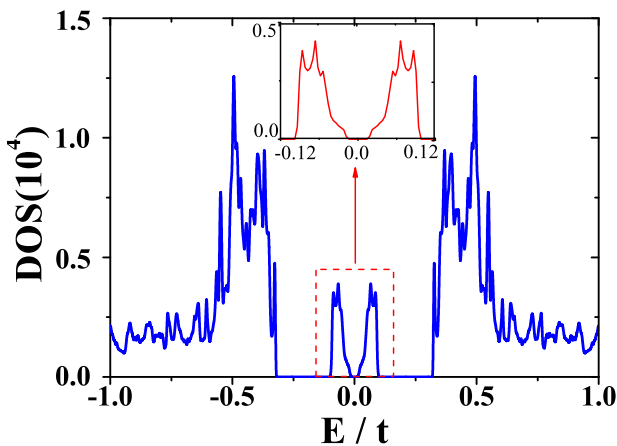}}
\caption{(Color online) The DOS of topological SC with square vortex-lattice:
the mid-gap states are induced by the vortex-lattice. The inset shows the
details of the mid-gap energy band. }%
\label{dos1}%
\end{figure}

\begin{figure}[ptb]
\scalebox{0.34}{\includegraphics[width = 25.0cm]{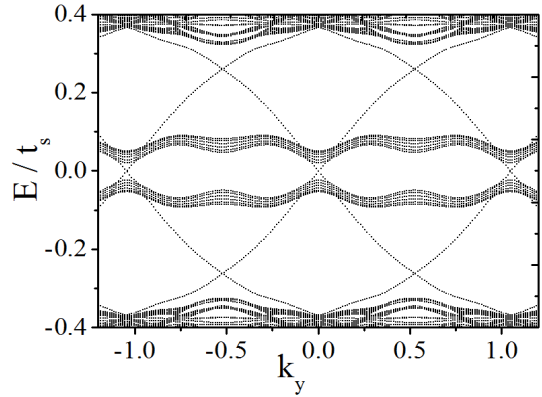}} \caption{(Color
online) The spectrum flow of topological SC with square vortices lattice on a
cylinder. There exist gapless edge states for the case of $u=-4t_{s}$,
$h=0.8t_{s}$, $\lambda=0.5t_{s}$, $\Delta=0.5t_{s}$.}%
\label{sfl}%
\end{figure}

\begin{figure}[ptb]
\scalebox{0.34}{\includegraphics[width = 25.0cm]{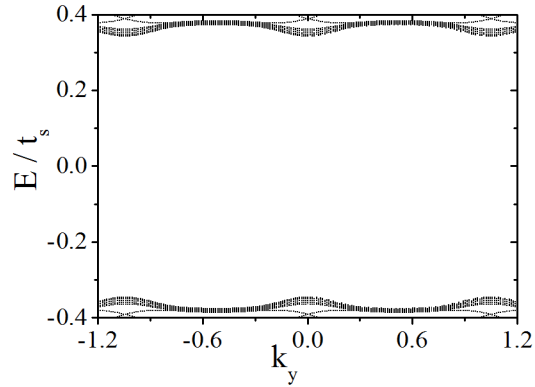}} \caption{(Color
online) The spectrum flow of normal SC with square vortex-lattice on a
cylinder. The parameters are $u=-4t_{s}$, $h=0.2t_{s}$, $\lambda=0.5t_{s}$,
$\Delta=0.5t_{s}$. There is no edge state. }%
\label{sf2}%
\end{figure}

The tunneling effect between vortices would lead to the coupling between two
Majorana fermions around the vortices. Thus there exists a mid-gap energy band
for the Majorana fermions. We study the density of state (DOS) in the
topological SC with a square vortex-lattice numerically and the results are
shown in Fig.\ref{dos1}. From Fig.\ref{dos1}, one can see that besides the
energy bands of the paired electrons there exists a mid-gap energy band in
parent topological SC. In particular, the mid-gap energy band has finite
energy gap. It means that this mid-gap system as shown in Fig.\ref{dos1} may
be a topological state associated with a non-trivial topological number
intuitively. To check the topological properties of the mid-gap system, we
calculate its edge states. We consider a system on a cylinder with $12$
super-unitcells along $x$-direction while periodic boundary along
$y$-direction. Thus, $k_{y}$ is still a good quantum number and permits the
Fourier transformation $c_{k_{y}}(j_{x})=\frac{1}{\sqrt{L_{y}}}\sum_{j_{y}%
}c(j_{x},j_{y})e^{ik_{y}j_{y}}.$ The spectral flow of this system on a
cylinder is plotted in Fig.\ref{sfl}. From Fig.\ref{sfl}, one can see that two
gapless chiral edge states are localized at the boundaries. On the other hand,
we also plot the spectral flow in Fig.\ref{sf2}. When the parent SC is a
non-topological SC, the edge states disappear. As a result, we conclude that
the mid-gap system shown in Fig.\ref{dos1} is indeed a topological state. From
Fig.9, we also observe the disappear of the mid-gap band which is induced by
the vortex-lattice. So the Majorana mode around the $\pi$-flux is protected by
topological invariant of the parent $s$-wave SC\textbf{.} 

\section{The topological Majorana lattice model}

In the last section we found a topological mid-gap system of the s-wave
topological SC with a square vortex-lattice numerically. In this section we
will learn the nature of this topological mid-gap system analytically.

\begin{figure}[ptb]
\begin{center}
\scalebox{0.9}{\includegraphics* [3.2in,3.8in][8in,5.8in]{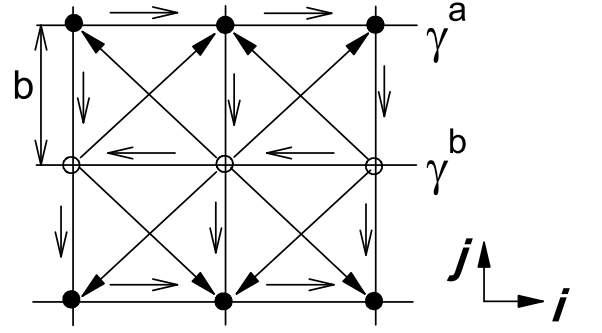}}
\end{center}
\caption{ Two sublattices of the Majorana lattice: $\gamma^{a}$ and
$\gamma^{b}$. The arrows denote the $e^{i\pi/2}$ of the hopping parameters.
Accordingly, there exists $\pi/2$ in each triangular plaquette. The lattice
constant of the Majorana lattice model is $b=6a$.}%
\label{gag}%
\end{figure}

We propose an effective description of the $s$-wave topological SC with a
square vortex-lattice, of which each vortex traps a Majorana zero mode and two
Majorana zero modes couple with each other by a short range interaction. The
interaction strength is just the energy splitting $\delta E$ from the
intervortex quantum tunneling. We call this effective description as the
Majorana lattice model, of which the tight-binding Hamiltonian can be written
as
\begin{equation}
\mathcal{H}_{m.f}=i\sum_{(j,k)}s_{jk}t_{jk}\gamma_{k}\gamma_{j}%
\end{equation}
where $t_{jk}$ is the hopping amplitude from $j$ to $k$, and satisfies
$t_{jk}=t_{jk}^{\ast}$. $\gamma_{j}$ is a Majorana operator ($\gamma_{j}%
^{\dag}=\gamma_{j}$) obeying anti-commutate relation $\{ \gamma_{j},\gamma
_{k}\}=2\delta_{jk}$. $s_{ij}=-s_{ji}$ is a gauge factor. The pair $(j,k)$
denotes the summation that runs over all the nearest neighbor (NN) pairs (with
hopping amplitude $t$) and all the next-nearest neighbor (NNN) pairs (with
hopping amplitude $t^{\prime}$). From the polygon rule proposed in
Ref.\cite{EGR}, each triangular plaquette possesses $\pi/2$ quantum flux
effectively. This Hamiltonian allows a $Z_{2}$ gauge choice $s_{jk}=\pm1$. For
our Majorana lattice model, one possible gauge is given by Fig.\ref{gag}.
Thus, the total number of Majorana modes $N$ must be even and then we can
divide the Majorana lattice into two sublattices.

Then the Majorana modes can be combined pairwise to create $N/2$ complex
fermionic states by pairing the Majorana operators as
\begin{equation}
\gamma_{2j-1}^{a}=a_{j}+a_{j}^{\dag},\quad \gamma_{2j}^{b}=(a_{j}-a_{j}^{\dag
})/i
\end{equation}
where $a_{j}$ ($a_{j}^{\dag}$) annihilates (creates) a fermion at link $j$. In
terms of operator $a_{j}$, the Majorana Hamiltonian takes the form of a
"\emph{topological SC}" state as follows:
\begin{align}
\mathcal{H}_{m.f}  &  =\sum_{\mathbf{x}}\big[t(a_{\mathbf{x}}^{\dag
}a_{\mathbf{x}+\mathbf{j}}-a_{\mathbf{x}}a_{\mathbf{x}+\mathbf{j}}^{\dag
})+t(a_{\mathbf{x}}a_{\mathbf{x}+\mathbf{j}}-a_{\mathbf{x}}^{\dag
}a_{\mathbf{x}+\mathbf{j}}^{\dag})\nonumber \\
&  +2t(a_{\mathbf{x}}^{\dag}a_{\mathbf{x}+\mathbf{i}}-a_{\mathbf{x}%
}a_{\mathbf{x}+\mathbf{i}}^{\dag})-2it(a_{\mathbf{x}}a_{\mathbf{x}+\mathbf{i}%
}+a_{\mathbf{x}}^{\dag}a_{\mathbf{x}+\mathbf{i}}^{\dag})\nonumber \\
&  +t^{\prime}(a_{\mathbf{x}}^{\dag}a_{\mathbf{x}+\mathbf{i}-\mathbf{j}%
}-a_{\mathbf{x}}a_{\mathbf{x}+\mathbf{i}-\mathbf{j}}^{\dag}-a_{\mathbf{x}%
}a_{\mathbf{x}+\mathbf{i}-\mathbf{j}}+a_{\mathbf{x}}^{\dag}a_{\mathbf{x}%
+\mathbf{i}-\mathbf{j}}^{\dag})\nonumber \\
&  +t^{\prime}(a_{\mathbf{x}}^{\dag}a_{\mathbf{x}+\mathbf{i}+\mathbf{j}%
}-a_{\mathbf{x}}a_{\mathbf{x}+\mathbf{i}+\mathbf{j}}^{\dag}+a_{\mathbf{x}%
}a_{\mathbf{x}+\mathbf{i}+\mathbf{j}}-a_{\mathbf{x}}^{\dag}a_{\mathbf{x}%
+\mathbf{i}+\mathbf{j}}^{\dag})\nonumber \\
&  -t(a_{\mathbf{x}}^{\dag}a_{\mathbf{x}}-a_{\mathbf{x}}a_{\mathbf{x}}^{\dag
})\big]
\end{align}
where $\mathbf{i}$ and $\mathbf{j}$ are two orthogonal unit vectors
(Fig.\ref{gag}). 

\begin{figure}[ptb]
\scalebox{0.33}{\includegraphics* [0in,0in][10in,7.2in]{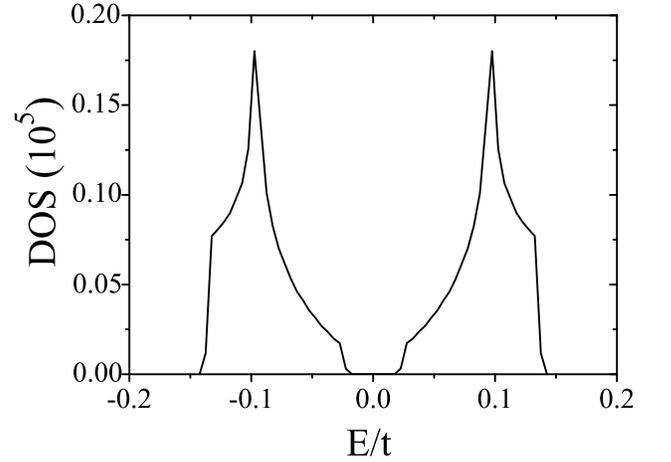}}
\caption{(Color online) The DOS of Majorana lattice model for the case of
$t=0.048t_{s}$ and $t^{\prime}=0.006t_{s}$. }%
\label{dos2}%
\end{figure}

Performing a Fourier transformation $a_{k}=\frac{1}{\sqrt{L_{\mathbf{x}}}}%
\sum_{\mathbf{x}}a_{\mathbf{x}}e^{-ikR_{\mathbf{x}}},$ the Hamiltonian
becomes
\begin{equation}
\mathcal{H}_{m.f}=\sum_{k,a}\psi_{k}^{\dag}d^{a}(k)\tau_{a}\psi_{k}%
\end{equation}
in the basis $\psi_{k}^{\dag}=(a_{k}^{\dag},a_{-k})$, where $a=x,y,z$,
$\tau_{a}$ is the Pauli matrix, and
\begin{align}%
\begin{split}
d^{z}(k)  &  =-2t\sin^{2}k_{y}+4t^{\prime}\cos k_{x}\cos^{2}k_{y}\\
d^{y}(k)  &  =-\sin2k_{y}(t+2t^{\prime}\cos k_{x})\\
d^{x}(k)  &  =-2t\sin k_{x}%
\end{split}
\end{align}
See the detailed calculations in Appendix. Then we get the energy spectrum
$E(k)$ of the Majorana lattice model as
\begin{equation}
E(k)=\pm \sqrt{\sum_{a=x,y,z}\left \vert d_{a}(k)\right \vert ^{2}}.
\end{equation}
From this result we can derive that there always exists an\textbf{ }energy gap
of the Majorana lattice model as long as $t^{\prime}\neq0$. Then we calculate
the DOS of the Majorana lattice model and show the result in Fig.\ref{dos2}.
One may compare the DOS of the Majorana lattice model in Fig.\ref{dos2} and
the DOS of the mid-gap system in Fig.\ref{dos1} and find the similarity
between them. 

\begin{figure}[ptb]
\scalebox{0.33}{\includegraphics* [0in,0in][10in,7.2in]{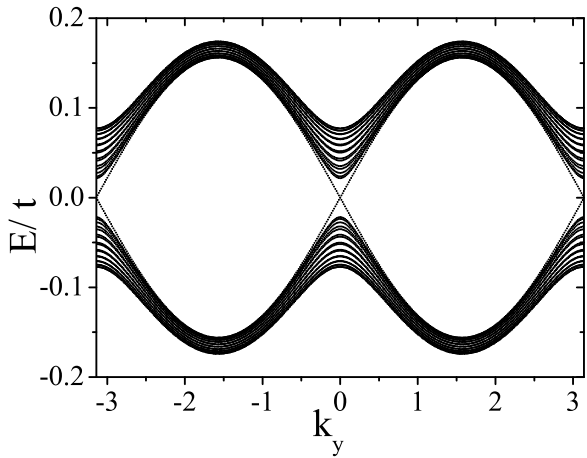}}
\caption{(Color online) The edge state of Majorana lattice model for the case
of $t=0.048t_{s}$, $t^{\prime}=0.006t_{s}.$}%
\label{edge2}%
\end{figure}

To characterize the topological properties of the Majorana lattice model, we
define the nontrivial topological invariant - the Chern number
\begin{equation}
\mathcal{Q}=\frac{1}{4\pi}\int \int_{BZ}d^{2}k\frac{1}{|\mathbf{d}(k)|^{3}%
}\mathbf{d}(k)\cdot \frac{\partial \mathbf{d}(k)}{\partial k_{x}}\times
\frac{\partial \mathbf{d}(k)}{\partial k_{y}}%
\end{equation}
which measures that the unit vector $\mathbf{d}(k)/|\mathbf{d}(k)|$ maps the
Brillouin zone boundary onto sphere $S^{2}$ via the Chern number $\mathcal{Q}%
$. In the presence of NNN hopping $t^{\prime}$, we have $\mathcal{Q}=\pm1$.
Furthermore, we calculate the edge states of the Majorana lattice model and
present the result in Fig.\ref{edge2}. One may compare the spectrum of the
edge states of the Majorana lattice model in Fig.\ref{edge2} with that of the
mid-gap system in Fig.\ref{sfl} and also find the similarity between them. Now
we can conclude that the (topological) Majorana lattice model captures the key
low energy physics of the topological SC with a square vortex-lattice.

In addition, we finish this section by a brief discussion of the $s$-wave
topological SC with triangle vortex-lattice. If in the region of $|t^{\prime
}/t|<<1$, we observe that the model is characterized by the winding number
$\mathcal{Q}=\pm1$, and this may stem from the redundant NN hopping (which
behave as the NNN hopping of the square Majorana lattice model) of the
triangular lattice. Moreover, the exotic phase $\mathcal{Q}=\pm3$ can be
reached by varying of the ratio $t^{\prime}/t$. In the region of $|t^{\prime
}/t|>>1$, the triangular Majorana lattice model has a $\mathcal{Q}=\pm3$
phase. It's worth to point out that, for the triangular Majorana lattice model
in $p+ip$ SC, or interacting vortex-lattice in Kitaev's honeycomb model,
people may find similar phase diagram\cite{VILL,p}.

\section{Conclusion}

In the end, we draw a conclusion. In this paper we studied the properties of
an $s$-wave topological SC with a square vortex-lattice. Because each vortex
traps a Majorana zero mode, when we took into account of zero mode tunneling
that couples the vortex sites, the Majorana zero modes of the vortex-lattice
form a Majorana lattice model. We found that this Majorana lattice model shows
nontrivial topological properties, including a nonvanishing Chern number,
chiral Majorana edge state. In this sense, the Majorana lattice model is
really a "topological SC" on the parent topological SC. And the "topological
SC" induced by the vortex lattice is topologically protected by the
topological invariant of the parent $s$-wave SC. Such correspondence between
the topological properties of the Majorana lattice model and the topological
properties of the $s$-wave topological SC model with vortex-lattice is another
holographic feature of the topological state. In addition, we also had used
the numerical approach to study the $s$-wave topological SC with a square
vortex-lattice and got similar results.

Furthermore, the Majorana lattice model for chiral $p+ip$ topological
superconductor with vortex-lattice and that for the coupled system due to the
proximity effect between $s$-wave SC and three dimensional topological
insulator with vortex-lattice have the same topological properties to our
case. Thus, this approach paves a new way to observe the Majorana modes for
the topological SC.

\begin{acknowledgments}
This work is supported by National Basic Research Program of China (973
Program) under the grant No. 2011CB921803, 2012CB921704 and NFSC Grant No.11174035.
\end{acknowledgments}

\appendix  

\section{Fourier transformation of the Hamiltonian for the Majorana lattice
model}

The Hamiltonian of the Majorana lattice model can be obtained by Fourier
transformation $a_{\mathbf{x}}=1/\sqrt{L_{\mathbf{x}}}\sum_{\mathbf{k}%
}a_{\mathbf{k}}e^{-i\mathbf{k}R_{\mathbf{x}}},$%
\begin{align}
\mathcal{H}_{m.f}  &  =1/L_{\mathbf{x}}\sum_{\mathbf{k},\mathbf{k^{\prime}}%
}\sum_{\mathbf{x}}\nonumber \\
&  \{t(a_{\mathbf{k}}^{\dag}a_{\mathbf{k^{\prime}}}e^{i\mathbf{k}%
R_{\mathbf{x}}-i\mathbf{k^{\prime}}R_{\mathbf{x}+\mathbf{j}}}-a_{\mathbf{k}%
}a_{\mathbf{k^{\prime}}}^{\dag}e^{-i\mathbf{k}R_{\mathbf{x}}%
+i\mathbf{k^{\prime}}R_{\mathbf{x}+\mathbf{j}}})\nonumber \\
&  +t(a_{\mathbf{k}}a_{\mathbf{k^{\prime}}}e^{-i\mathbf{k}R_{\mathbf{x}%
}-i\mathbf{k^{\prime}}R_{\mathbf{x}+\mathbf{j}}}-a_{\mathbf{k}}^{\dag
}a_{\mathbf{k^{\prime}}}^{\dag}e^{i\mathbf{k}R_{\mathbf{x}}+i\mathbf{k^{\prime
}}R_{\mathbf{x}+\mathbf{j}}})\nonumber \\
&  +2t^{\prime}(a_{\mathbf{k}}^{\dag}a_{\mathbf{k^{\prime}}}e^{i\mathbf{k}%
R_{\mathbf{x}}-i\mathbf{k^{\prime}}R_{\mathbf{x}+\mathbf{i}}}-a_{\mathbf{k}%
}a_{\mathbf{k^{\prime}}}^{\dag}e^{-i\mathbf{k}R_{\mathbf{x}}%
+i\mathbf{k^{\prime}}R_{\mathbf{x}+\mathbf{i}}})\nonumber \\
&  -2it(a_{\mathbf{k}}a_{\mathbf{k^{\prime}}}e^{-i\mathbf{k}R_{\mathbf{x}%
}-i\mathbf{k^{\prime}}R_{\mathbf{x}+\mathbf{i}}}+a_{\mathbf{k}}^{\dag
}a_{\mathbf{k^{\prime}}}^{\dag}e^{i\mathbf{k}R_{\mathbf{x}}+i\mathbf{k^{\prime
}}R_{\mathbf{x}+\mathbf{i}}})\nonumber \\
&  +t^{\prime}(a_{\mathbf{k}}^{\dag}a_{\mathbf{k^{\prime}}}e^{i\mathbf{k}%
R_{\mathbf{x}}-i\mathbf{k^{\prime}}R_{\mathbf{x}+\mathbf{i}-\mathbf{j}}%
}-a_{\mathbf{k}}a_{\mathbf{k^{\prime}}}^{\dag}e^{-i\mathbf{k}R_{\mathbf{x}%
}+i\mathbf{k^{\prime}}R_{\mathbf{x}+\mathbf{i}-\mathbf{j}}})\nonumber \\
&  -t^{\prime}(a_{\mathbf{k}}a_{\mathbf{k^{\prime}}}e^{-i\mathbf{k}%
R_{\mathbf{x}}-i\mathbf{k^{\prime}}R_{\mathbf{x}+\mathbf{i}-\mathbf{j}}%
}-a_{\mathbf{k}}^{\dag}a_{\mathbf{k^{\prime}}}^{\dag}e^{i\mathbf{k}%
R_{\mathbf{x}}+i\mathbf{k^{\prime}}R_{\mathbf{x}+\mathbf{i}-\mathbf{j}}%
})\nonumber \\
&  +t^{\prime}(a_{\mathbf{k}}^{\dag}a_{\mathbf{k^{\prime}}}e^{i\mathbf{k}%
R_{\mathbf{x}}-i\mathbf{k^{\prime}}R_{\mathbf{x}+\mathbf{i}+\mathbf{j}}%
}-a_{\mathbf{k}}a_{\mathbf{k^{\prime}}}^{\dag}e^{-i\mathbf{k}R_{\mathbf{x}%
}+i\mathbf{k^{\prime}}R_{\mathbf{x}+\mathbf{i}+\mathbf{j}}})\nonumber \\
&  +t^{\prime}(a_{\mathbf{k}}a_{\mathbf{k^{\prime}}}e^{-i\mathbf{k}%
R_{\mathbf{x}}-i\mathbf{k^{\prime}}R_{\mathbf{x}+\mathbf{i}+\mathbf{j}}%
}-a_{\mathbf{k}}^{\dag}a_{\mathbf{k^{\prime}}}^{\dag}e^{i\mathbf{k}%
R_{\mathbf{x}}+i\mathbf{k^{\prime}}R_{\mathbf{x}+\mathbf{i}+\mathbf{j}}%
})\nonumber \\
&  -t(a_{\mathbf{k}}^{\dag}a_{\mathbf{k^{\prime}}}e^{i\mathbf{k}R_{\mathbf{x}%
}-i\mathbf{k^{\prime}}R_{\mathbf{x}}}-a_{\mathbf{k}}a_{\mathbf{k^{\prime}}%
}^{\dag}e^{-i\mathbf{k}R_{\mathbf{x}}+i\mathbf{k^{\prime}}R_{\mathbf{x}}})\}
\end{align}
where $R_{\mathbf{x}+\mathbf{\delta}}=R_{\mathbf{x}}+\mathbf{\delta}$,
($\mathbf{\delta}=\mathbf{i}, \mathbf{j}, \mathbf{i}+\mathbf{j}$). Using the
identity $\delta_{\mathbf{k},\mathbf{k}^{\prime}}=1/L_{\mathbf{x}}%
\sum_{\mathbf{x}}e^{i(\mathbf{k}-\mathbf{k}^{\prime})R_{\mathbf{x}}}$, we
have
\begin{align}
\mathcal{H}_{m.f}  &  =\sum_{\mathbf{k}}\{t(a_{k}^{\dag}a_{k}e^{-ik_{y}}%
-a_{k}a_{k}^{\dag}e^{ik_{y}})\nonumber \\
&  +t(a_{k}a_{-k}e^{ik_{y}}-a_{k}^{\dag}a_{-k}^{\dag}e^{-ik_{y}})\nonumber \\
&  +2t^{\prime}(a_{k}^{\dag}a_{k}e^{-ik_{x}}-a_{k}a_{k}^{\dag}e^{ik_{x}%
})\nonumber \\
&  -2it(a_{k}a_{-k}e^{ik_{x}}+a_{k}^{\dag}a_{-k}^{\dag}e^{-ik_{x}})\nonumber \\
&  +t^{\prime}(a_{k}^{\dag}a_{k}e^{-i(k_{x}-k_{y})}-a_{k}a_{k}^{\dag
}e^{i(k_{x}-k_{y})})\nonumber \\
&  -t^{\prime}(a_{k}a_{-k}e^{i(k_{x}-k_{y})}-a_{k}^{\dag}a_{-k}^{\dag
}e^{-i(k_{x}-k_{y})})\nonumber \\
&  +t^{\prime}(a_{k}^{\dag}a_{k}e^{-i(k_{x}+k_{y})}-a_{k}a_{k}^{\dag
}e^{i(k_{x}+k_{y})})\nonumber \\
&  +t^{\prime}(a_{k}a_{-k}e^{i(k_{x}+k_{y})}-a_{k}^{\dag}a_{-k}^{\dag
}e^{-i(k_{x}+k_{y})})\nonumber \\
&  -t(a_{k}^{\dag}a_{k}-a_{k}a_{k}^{\dag})\}
\end{align}
which is
\begin{align}
\mathcal{H}_{m.f}  &  =(t\cos k_{y}+2t^{\prime}\cos k_{x}+2t^{\prime}\cos
k_{x}\cos k_{y})a_{k}^{\dag}a_{k}\nonumber \\
&  -(t\cos k_{y}+2t^{\prime}\cos k_{x}+2t^{\prime}\cos k_{x}\cos k_{y}%
)a_{k}a_{k}^{\dag}\nonumber \\
&  -(2t\sin k_{x}+it\sin k_{y}+2it^{\prime}\sin k_{y}\cos k_{x})a_{-k}%
a_{k}\nonumber \\
&  -(2t\sin k_{x}-it\sin k_{y}-2it^{\prime}\sin k_{y}\cos k_{x})a_{k}^{\dag
}a_{-k}^{\dag}%
\end{align}
or
\begin{equation}
\mathcal{H}_{m.f}=\sum_{\mathbf{k}}\sum_{a=x,y,z}\psi_{\mathbf{k}}^{\dag}%
d^{a}(\mathbf{k})\sigma_{a}\psi_{\mathbf{k}}.
\end{equation}
where $\psi_{\mathbf{k}}^{\dag}=(a_{\mathbf{k}}^{\dag},a_{-\mathbf{k}}).$


\begin{thebibliography}{99}                                                                                               %


\bibitem {YT}Y. Tsutsumi, T. Kawakami, T. Mizushima, M. Ichioka, and K.
Machida, Phys. Rev. Lett. \textbf{101}, 135302 (2008).

\bibitem {SDS}S. Das Sarma, C. Nayak, and S. Tewari, Phys. Rev. B \textbf{73},
220502R (2006).

\bibitem {LFu}L. Fu and C. L. Kane, Phys. Rev. Lett. \textbf{100}, 096407 (2008).

\bibitem {NR}N. Read and D. Green, Phys. Rev. B \textbf{61}, 10267 (2000).

\bibitem {MS}M. Sato, Y. Takahashi, and S. Fujimoto, Phys. Rev. Lett.
\textbf{103}, 020401 (2009).

\bibitem {JDS}J. D. Sau, R. M. Lutchyn, S. Tewari, and S. Das Sarma, Phys.
Rev. Lett. \textbf{104}, 040502 (2010).

\bibitem {RML}R. M. Lutchyn, J. D. Sau, and S. Das Darma, Phys. Rev. Lett.
\textbf{105}, 077001 (2010).

\bibitem {DAI}D. A. Ivlovik, Phys. Rev. Lett. \textbf{86}, 268 (2001).

\bibitem {IPR}I. P. Radu, J. B. Miller, C. M. Marcus, M.A.Kastner, L. N.
Pfeiffer, K. W. West, Science \textbf{320}, 899 (2008).

\bibitem {WBI}W. Bishara, P. Bonderson, C. Nayak, K. Shtengel, J. K.
Slingerland, Phys. Rev. B \textbf{80}, 155303 (2009).

\bibitem {JAL}J. Alicea, Phys. Rev. B \textbf{81}, 125318 (2010).

\bibitem {CNA}C. Nayak, S. H. Simon, A. Stern, M. Freedman, and S. Das Sarma,
Rev. Mod. Phys. \textbf{80}, 1083 (2008).

\bibitem {SDAS}S. Das Sarma, M. Freedman, and C. Nayak. Phys. Rev. Lett.
\textbf{94}, 166802 (2005).

\bibitem {STE}S. Tewari, S. Das Sarma, C. Nayak, C. Zhang, and P. Zoller,
Phys. Rev. Lett. \textbf{98}, 010506 (2007).

\bibitem {AKI}A. Kitaev, Ann. Phys. \textbf{303}, 2 (2003). A. Kitaev, Ann.
Phys. \textbf{321}, 2 (2006).

\bibitem {CG}C. Gils, E.Ardonne, S. Trebst, A. W. W. Ludwig, M. Troyer, and Z.
Wang, Phys. Rev. Lett. \textbf{103}, 070401 (2009).

\bibitem {AF}A. Feiguin, S. Trebst, A. W. W. Ludwig, M. Troyer, A. Kitaev, Z.
Wang, and M. H. Freedman, Phys. Rev. Lett. \textbf{98}, 160409 (2007).

\bibitem {MCH2}M. Cheng, R. M. Lutchyn, V. Galitski, and S. Das Sarma, Phys.
Rev. B \textbf{82}, 094504 (2010).

\bibitem {MCH}M. Cheng, R. M. Lutchyn, V. Galitski, and S. Das Sarma, Phys.
Rev. Lett \textbf{103}, 107001 (2009).

\bibitem {ASTE}A. Stern, F. von Oppen, and E. Mariani, Phys. Rev. B
\textbf{70}, 205338 (2004).

\bibitem {VGU}V. Gurarie, and L. Radzihovsky, Ann. Phys. \textbf{322}, 2 (2007).

\bibitem {TMI}T. Mizushima, M. Ichioka, and K. Machida, Phys. Rev. Lett.
\textbf{101}, 150409 (2008).

\bibitem {CZH}C. Zhang, S. Tewari, R. M. Lutchyn, and S. Das Sarma, Phys. Rev.
Lett. \textbf{101}, 160401 (2008).

\bibitem {NRC}N. R. Cooper, and G. V. Shiyapnikov, Phys. Rev. Lett.
\textbf{103},155302 (2009).

\bibitem {YNS}Y. Nishida, Ann. Phys. (N.Y.) \textbf{324}, 897 (2009).

\bibitem {VILL}V. Lahtinen, A. W. Ludwig, J. K.Pachos, and S. Trebst, ArXiv:1111.3296.

\bibitem {MBA}M. Baraban, G. Zikos, N. Bonesteel, and S. H. Simon, ArXiv:0901.3502.

\bibitem {EGR}E. Grosfeld and Ady Stern, Phys. Rev. \textbf{B 73}, 201303 (2006).

\bibitem {p}J. K. Pachos, E. Alba, V. Lahtinen, J. J. Garcia-Ripoll, ArXiv:1209.5115.
\end{thebibliography}
\end{document}